\documentclass[aps,superscriptaddress,twocolumn,dvips]{revtex4}
\usepackage{feynmp}
\usepackage{amsmath}
\usepackage{amssymb}
\usepackage{epsfig}
\usepackage{graphicx}
\usepackage{subfigure}
\usepackage{hyperref}
\usepackage{bbm}
\usepackage{color}
\usepackage{longtable}
\usepackage{booktabs}
\usepackage{multirow}
\usepackage{array}
\newcommand{\PreserveBackslash}[1]{\let\temp=\\#1\let\\=\temp}
\newcolumntype{C}[1]{>{\PreserveBackslash\centering}p{#1}}
\newcolumntype{R}[1]{>{\PreserveBackslash\raggedleft}p{#1}}
\newcolumntype{L}[1]{>{\PreserveBackslash\raggedright}p{#1}}

\newcommand{\Rmnum}[1]{\expandafter\@slowromancap\romannumeral #1@}
\allowdisplaybreaks

\begin{document}

\title{Excitonic pairing of two-dimensional Dirac fermions near
the antiferromagnetic quantum critical point}

\author{Hai-Xiao Xiao}
\affiliation{Department of Modern Physics, University of Science and
Technology of China, Hefei, Anhui 230026, P. R. China}

\author{Jing-Rong Wang}
\affiliation{Anhui Province Key Laboratory of Condensed Matter
Physics at Extreme Conditions, High Magnetic Field Laboratory of the
Chinese Academy of Science, Hefei, Anhui 230031, P. R. China}

\author{Zheng-Wei Wu}
\affiliation{Department of Modern Physics, University of Science and
Technology of China, Hefei, Anhui 230026, P. R. China}

\author{Guo-Zhu Liu}
\altaffiliation{gzliu@ustc.edu.cn} \affiliation{Department of Modern
Physics, University of Science and Technology of China, Hefei, Anhui
230026, P. R. China}

\begin{abstract}
Two-dimensional Dirac fermions are subjected to two types of
interactions, namely the long-range Coulomb interaction and the
short-range on-site interaction. The former induces excitonic
pairing if its strength $\alpha$ is larger than some critical value
$\alpha_c$, whereas the latter drives an antiferromagnetic Mott
transition when its strength $U$ exceeds a threshold $U_c$. Here, we
study the impacts of the interplay of these two interactions on
excitonic pairing with the Dyson-Schwinger equation approach. We
find that the critical value $\alpha_c$ is increased by weak
short-range interaction. As $U$ increases to approach $U_c$, the
quantum fluctuation of antiferromagnetic order parameter becomes
important and interacts with the Dirac fermions via the Yukawa
coupling. After treating the Coulomb interaction and Yukawa coupling
interaction on an equal footing, we show that $\alpha_c$ is
substantially increased as $U \rightarrow U_c$. Thus, the excitonic
pairing is strongly suppressed near the antiferromagnetic quantum
critical point. We obtain a global phase diagram on the $U$-$\alpha$
plane, and illustrate that the excitonic insulating and
antiferromagnetic phases are separated by an intermediate semimetal
phase. These results provide a possible explanation of the
discrepancy between recent theoretical progress on excitonic gap
generation and existing experiments in suspended graphene.
\end{abstract}

\maketitle

\section{INTRODUCTION\label{Sec:Intro}}

Two-dimensional (2D) massless Dirac fermions are the low-energy
excitations of a number of condensed matter systems. Examples
include $d$-wave high-$T_c$ cuprate superconductors \cite{LeeRMP,
Orenstein}, graphene \cite{Novoselov04, Novoselov05, CastroNeto,
DasSarma, Kotov}, surface of three-dimensional (3D) topological
insulators \cite{topo}, and organic conductor
$\alpha$-(BEDT-TTF)$_2$I$_3$ \cite{Hirata16}. While the single
particle properties of Dirac fermion systems have already been
extensively studied, the strong correlation effects are still not
well understood. Ordinary metals are known to be robust against
repulsive interactions \cite{Shankar94}, which renders the validity
of Fermi liquid theory. In contrast, the repulsive interactions are
much more important in 2D Dirac fermion systems, and may lead to
several possible phase-transition instabilities \cite{CastroNeto,
DasSarma, Kotov, Downing19}. Generically, there are two types of
repulsive interactions, namely long-range Coulomb interaction and
Hubbard-like on-site interaction. The former is spin blinded,
whereas the latter acts on two electrons with different spins and is
thus spin distinguished.

When the strength parameter $U$ of on-site repulsive interaction is
greater than a critical value $U_c$, there is a quantum phase
transition from gapless semimetal (SM) to antiferromagnetic (AFM)
Mott insulator \cite{Herbut06}. The SM-AFM quantum critical point
(QCP) falls in the university class of Gross-Neveu-Yukawa model
\cite{Herbut06}. Apart from SM-AFM transition, Sato \emph{et al.}
\cite{Sato17} studied the transition between SM and Kekul\'{e}
valence-bond solid caused by on-site interaction. When other sorts
of repulsion are considered, SM materials could exhibit richer
phase-transition structures \cite{Raghu08, Weeks10, Grushin13,
Daghofer14, Sato17, Zeng18, Tang18, Buividovich18}. For instance,
Raghu \emph{et al.} \cite{Raghu08} investigated the cooperative
effects of nearest- and next-nearest- neighbor
repulsions, and found a number of insulating phases, including
charge density wave (CDW), AFM, and topological Mott phases that
display quantum anomalous Hall (QAH) and quantum spin Hall (QSH)
effects, although subsequent studies revealed that the topological
Mott phases can be destroyed by fluctuations \cite{Daghofer14}.

In case the Fermi level is located exactly at the band-touching
point, the long-range Coulomb interaction is poorly screened due to
the vanishing of density of states (DOS). If the Coulomb interaction
is weak, the system remains gapless, but the fermion velocity is
substantially renormalized \cite{Kotov, Geim11}. When the Coulomb
interaction strength parameter $\alpha$ exceeds a critical value
$\alpha_c$, a finite energy gap is dynamically generated via the
formation of excitonic-type particle-hole pairs
\cite{Khveshchenko01, Gorbar02, Khveshchenko04, Liu09,
Khveshchenko09, Gamayun09, Sabio10, Gamayun10, WangLiu12,
Popovici13, Gonzalez15, Carrington16, Carrington18, Xiao17, Drut09A,
Drut09B, Drut09C, Armour10, Armour11, Buividovich12, Ulybyshev13,
Smith14, Juan12, Kotikov16, Tupitsyn17}. This then turns the
originally gapless SM into a gapped excitonic insulator (EI).
Another interesting possibility is that the Coulomb-like interaction
can induce an electron-electron pairing, as predicted and discussed
in Refs.~\cite{Marnham15, Downing17}.

In previous works, the Coulomb interaction and the on-site
interaction were usually investigated separately. Their interplay
can give rise to intriguing properties, especially in the strong
interaction regimes. Interesting progress has recently been made
towards more detailed knowledge of this interplay. Tang \emph{et
al.} \cite{Tang18} have studied the influences of long-range Coulomb
and on-site interactions on the ground-state properties of 2D Dirac
fermion systems by combining the non-perturbative quantum Monte
Carlo (QMC) simulation and the renormalization group (RG) technique.
Their work \cite{Tang18} reproduced the previously discovered
logrithmic velocity renormalization and confirmed that SM-AFM
transition occurs at some critical value $U_c$. They further found
that $U_c$ increases as $\alpha$ grows, which indicates that the
Coulomb interaction disfavors AFM transition. These results are
summarized in the phase diagram shown in Figure~1 of
Ref.~\cite{Tang18}.

The results reported in Ref.~\cite{Tang18} are only applicable to
the region of weak Coulomb interaction. The region of strong Coulomb
interaction appears to be inaccessible to the numerical methods
developed in Ref.~\cite{Tang18} and Ref.~\cite{Buividovich18}. As
aforementioned, strong Coulomb interaction is able to induce
excitonic pairing and SM-EI transition. This problem has attracted
broad interest in the past two decades. Extensive theoretical
efforts have been devoted to examining whether the SM-EI transition
can be realized in graphene. In Refs.~\cite{Tang18, Buividovich18},
the influence of on-site interaction on SM-EI transition has not
been addressed. Moreover, it remains unclear how the SM-EI
transition is affected by the SM-AFM quantum criticality.

In this paper, we study the excitonic pairing of 2D Dirac fermions
by considering both the long-range Coulomb and on-site interactions.
In particular, we will investigate the impact of on-site interaction
on the fate of excitonic pairing. For small values of $U$, the
Coulomb interaction and on-site interaction need to be treated on an
equal footing. When $U$ grows, the AFM correlation is gradually
enhanced. As $U \rightarrow U_c$, the system approaches the AFM
QCP and the quantum fluctuations of AFM order parameter interacts
strongly with the Dirac fermions via the Yukawa-type coupling. To
examine the influence of AFM quantum criticality on excitonic
pairing, we need to study the interplay between the Coulomb
interaction and the Yukawa coupling interaction.

The non-perturbative Dyson-Schwinger (DS) equation approach will be
employed to compute the excitonic gap and to determine $\alpha_c$.
In our calculations, the series expansion is controlled by the small
parameter $1/N$, where $N$ is the spin degeneracy of Dirac fermion.
Within this framework, the Coulomb interaction parameter $\alpha$
can take any value. This allows us to access the strong Coulomb
interaction regime. The Yukawa coupling can also be handled by the
$1/N$ expansion. However, the on-site interaction is spin
distinguished, to be explained below, and the $1/N$ expansion
becomes invalid. In the case of weak on-site interaction, we will
perform weak coupling expansion.

After incorporating the impact of weak on-site interaction, we find
that the critical value $\alpha_c$ for EI transition is slightly
increased. At the AFM QCP ($U_c$), the value of $\alpha_c$ is
increased dramatically by the Yukawa coupling interaction. Indeed,
$\alpha_c$ is an increasing function of Yukawa coupling constant
$\lambda$. Apparently, excitonic pairing is significantly suppressed
by the quantum fluctuation of AFM order parameter. As $U$ decreases
from $U_c$, the system departs from AFM QCP and the suppression of
excitonic pairing caused by AFM fluctuation is weakened. Combining
these results with those reported in Ref.~\cite{Tang18}, we obtain a
schematic global phase diagram on the $U$-$\alpha$ plane, shown in
Fig.~1. It seems that the EI phase cannot be directly converted into
AFM Mott insulating phase: they are separated by an intermediate SM
phase.

\begin{figure}
\includegraphics[width=0.43\textwidth]{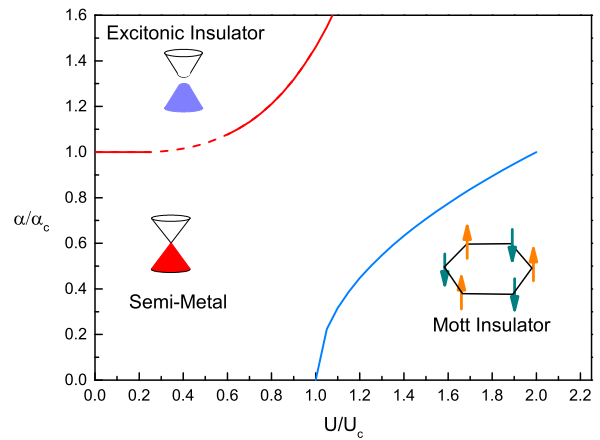}
\caption{The global phase diagram of 2D Dirac fermion system on the
plane spanned by Coulomb interaction parameter $\alpha$ and on-site
interaction parameter $U$. The critical line of $U_c$ is taken from
Ref.~\cite{Tang18}. The solid part of the EI critical line is
plotted based on our DS equation results, and the dashed part of
this line is plotted based on extrapolation.}
\label{Fig:ExcitonicMottphasediagram}
\end{figure}

Our theoretical results provide a possible explanation of the
discrepancy between recent theoretical progress and existing
experiments in graphene. It is known that $\alpha$ takes its maximal
value $\alpha = 2.16$ when graphene is suspended in vacuum. The
zero-temperature ground state of suspended graphene should be an
insulator if $\alpha_c < 2.16$. In a recent work, Carrington
\emph{et al.} \cite{Carrington18} has performed a careful DS
equation study by going beyond many of the previously used
approximations, and found that $\alpha_c \approx 2.0$, which is
slightly below $\alpha = 2.16$. This result suggests that suspendend
graphene would be insulating at zero temperature. However, this is
apparently at odds with previous experiments \cite{Geim11,
Mayorov12}. According to the analysis of Refs.~\cite{Wehling11},
graphene seems to be close to the AFM QCP, thus the impact of AFM
quantum criticality on $\alpha_c$ needs to be seriously taken into
account. Our results show that the proximity to AFM QCP
substantially increases the critical value $\alpha_c$, which makes
SM-EI transition very unlikely in realistic graphene.

The rest of the paper is organized as follows. In
Sec.~\ref{Sec:Model}, we present the DS equation for dynamical
excitonic gap. The gap equation is solved and analyzed in
Sec.~\ref{Sec:NumResults}, and the physical application of the
result is discussed in Sec.~IV. The results are summarized in
Sec.~\ref{Sec:Summary}.

\section{Dyson-Schwinger gap equation\label{Sec:Model}}

The free 2D Dirac fermions are described by the Lagrangian in
Minkowski space
\begin{eqnarray}
\mathcal{L}_{0} = \sum_{\sigma} \bar{\Psi}_{\sigma}(\tau,
\mathbf{x})i(\gamma_{0}\partial_{0} - v\gamma_{i}\partial_{i})
\Psi_{\sigma}(\tau,\mathbf{x}),
\end{eqnarray}
where $\Psi_{\sigma}$ is a four-component spinor field and
$\bar{\Psi}_{\sigma} = \Psi_{\sigma}^{\dagger}\gamma_0$. The index
$\sigma$ sums from $1$ to $N$, with $N=2$ being the spin degeneracy
of Dirac fermion. The $4 \times 4$ gamma matrices are defined via
Pauli matrices as $\gamma_{0,i} = \tau_{3}
\otimes(\sigma_{3},i\sigma_{2},-i\sigma_{1})$, which satisfy the
Clifford algebra. The fermion velocity $v$ is taken be a constant.

We will consider three different sorts of interactions, including
the long-range Coulomb interaction, the spinful on-site interaction,
as well as the Yukawa coupling between Dirac fermions and AFM
quantum fluctuation. If the system is far from the AFM QCP, we only
need to study the first two interactions. But when the system is
sufficiently close to the AFM QCP, the interplay of Coulomb
interaction and Yukawa coupling should be carefully investigated.
Below we present the effective field theories for these three
interactions in order.

\subsection{Pure Coulomb interaction}

The pure Coulomb interaction can be modeled by the following
Lagrangian
\begin{eqnarray}
\mathcal{L}_{C} = -ea_{0}\sum_{\sigma}\bar{\Psi}_{\sigma}
\gamma_{0}\Psi_{\sigma} + a_{0}
\frac{|\bigtriangledown|}{2e^{2}}a_{0},
\end{eqnarray}
where $a_{0}$ is an auxiliary scalar field introduced to represent
Coulomb interaction. It is easy to verify that the Lagrangian
$\mathcal{L}_{0} + \mathcal{L}_{C}$ respects the continuous chiral
symmetry $\Psi_{\sigma}\rightarrow
e^{i\gamma_{5}\theta}\Psi_{\sigma}$, where $\gamma_5 = -\sigma_2
\otimes \sigma_0$ anti-commutes with $\gamma_{0,i}$.

\begin{figure}
\includegraphics[width=0.4\textwidth]{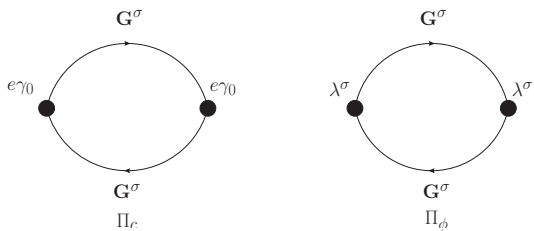}
\caption{Feynman diagrams of $\Pi_{c}$ and $\Pi_{\phi}$. The
difference between two diagrams lies in the expression of the
vertices.} \label{Fig:Coulomb polarization}
\end{figure}

The pure long-range Coulomb interaction has already been widely
studied \cite{Khveshchenko01, Gorbar02, Khveshchenko04, Liu09,
Khveshchenko09, Gamayun09, Sabio10, Gamayun10, WangLiu12,
Popovici13, Gonzalez15, Carrington16, Carrington18, Xiao17, Drut09A,
Drut09B, Drut09C, Armour10, Armour11, Buividovich12, Ulybyshev13,
Smith14, Juan12, Kotikov16, Tupitsyn17}. In Ref.~\cite{Downing17,
Downing19}, Downing and Portnoi have considered the problem of
electrostatic confinement of Dirac fermions and found zero-energy
bielectron bound state in scalar potentials. In this paper, we only
study the excitonic particle-hole pairing realized in systems with a
large number of Dirac fermions.

As mentioned in Sec.~\ref{Sec:Intro}, the Coulomb interaction can be
studied within the frame of $1/N$ expansion. Below, we will adopt an
approximation that retains only the leading order contribution of
$1/N$ expansion. To this order, the contribution of wave function
renormalization can be ignored. The free propagator of Dirac
fermions is
\begin{eqnarray}
G^{0}_{\sigma}(\omega,\mathbf{k}) = \frac{1}{\omega\gamma_{0} -
v(\gamma_{1}k_{x}+\gamma_{2}k_{y})}.
\end{eqnarray}
Interaction turns this free propagator into
\begin{eqnarray}
G_{\sigma}(\omega,\mathbf{k}) = \frac{1}{\omega\gamma_{0} -
v(\gamma_{1}k_{x}+\gamma_{2}k_{y}) -
m_{\sigma}(\omega,\mathbf{k})},
\end{eqnarray}
where $m_{\sigma}(\omega,\mathbf{k})$ is the fermion mass function.
Once $m_{\sigma}(\omega,\mathbf{k})$ acquires a finite value due to
the Coulomb interaction, an excitonic mass gap is generated and the
gapless SM is converted into a fully gapped EI. In order to examine
whether an excitonic gap is generated, we write down the following
DS equation
\begin{eqnarray}
m_{\sigma}(\varepsilon,\mathbf{p}) =\frac{1}{4} \int \frac{d\omega}{2
\pi}\frac{d^2\mathbf{k}}{(2\pi)^2} \textmd{Tr}[\gamma_{0}G_{\sigma}(\omega,
\mathbf{k})\gamma_{0}V(\Omega,\mathbf{q})],
\end{eqnarray}
where $\Omega=\varepsilon-\omega$ and $\mathbf{q} =
\mathbf{p}-\mathbf{k}$. Here, the effective Coulomb interaction is
given by
\begin{eqnarray}
V(\Omega,\mathbf{q}) = \frac{1}{V_{0}^{-1}(\Omega,\mathbf{q}) +
\Pi_{c}(\Omega,\mathbf{q})},\label{Eq:CoulombPropagator}
\end{eqnarray}
where $\Pi_{c}(\Omega,\mathbf{q})$ is the polarization function and
\begin{eqnarray}
V_{0}(\mathbf{q}) = \frac{2\pi e^{2}\delta(t)}{\kappa|\mathbf{q}|}
\end{eqnarray}
is the bare Coulomb interaction, with $\kappa = \epsilon_{0}
\epsilon_{r}$ being the dielectric constant. To the leading order of
$1/N$ expansion, the Feynman diagram for
$\Pi_{c}(\Omega,\mathbf{q})$ is shown in Fig.~\ref{Fig:Coulomb
polarization}. At the random phase approximation (RPA) level, the
one-loop $\Pi_{c}$ is calculated as follows \cite{Khveshchenko01}
\begin{eqnarray}
\Pi_{c}(\Omega,\mathbf{q}) &=& -N\int\frac{d\omega}{2\pi}
\frac{d^{2}\mathbf{k}}{(2\pi)^2} \textmd{Tr}[\gamma_{0}
G_{0}(\omega,\mathbf{k})\gamma_{0}\nonumber \\
&& G_{0}(\omega + \Omega,\mathbf{k}+\mathbf{q})] \nonumber \\
&=& \frac{N}{8}\frac{\mathbf{q}^{2}}{\sqrt{v\mathbf{q}^{2}-
\Omega^{2}}}.\label{Eq:CoulombPolarization}
\end{eqnarray}
After preforming the Wick rotation($\omega\rightarrow i\omega$), we
get the following DS gap equation in Euclidean space
\cite{Khveshchenko01}
\begin{eqnarray}
m_{\sigma}(\varepsilon,\mathbf{p}) &=& \int \frac{d\omega}{2
\pi}\frac{d^2\mathbf{k}}{(2\pi)^2} \frac{m_{\sigma}(\omega,
\mathbf{k})}{\omega^{2}+p_{x}^{2}+p_{y}^{2} +
m_{\sigma}(\omega,\mathbf{k})^{2}} \nonumber \\
&& \times \frac{1}{\frac{|\mathbf{q}|}{2\pi v\alpha} + \frac{N}{8}
\frac{\mathbf{q}^{2}}{\sqrt{\Omega^{2}+v^{2}\mathbf{q}^{2}}}}.
\label{Eq:CoulombgapEquation}
\end{eqnarray}
where a new parameter $\alpha = e^{2}/v\kappa$ is defined to measure
the effective interaction strength. For a given flavor $N$, the
above gap equation has a nontrivial solution, i.e., $m \neq 0$, only
when $\alpha > \alpha_c$. The QCP between SM and EI phases is
located at $\alpha = \alpha_c$. If the value of $\alpha$ is fixed, a
nonzero gap could be generated only when $N < N_c$.

\subsection{Weak on-site interaction\label{Sec:Model:Fourfermion interaction}}

As shown by Herbut and his collaborators \cite{Herbut06, Herbut09},
the generic on-site interaction is complex and can be decomposed
into eight independent four-fermion coupling terms. Here, following
Ref.~\cite{Tang18}, we only consider the spin-distinguished
interaction term
\begin{eqnarray}
\mathcal{L}_{I} = g\sum_{\sigma}(\sigma\bar{\Psi}_{\sigma}
\Psi_{\sigma})^{2},
\end{eqnarray}
which is responsible for the transition into AFM Mott insulating
phase. In Ref.~\cite{Tang18}, this is referred to as spinful
Gross-Neveu (GN) interaction. It is also called chiral Heisenberg GN
interaction \cite{Gracey18}. According to Ref.~\cite{Herbut06}, $g$
is related to $U$ through the identity $g = -\frac{Ua^{2}}{8}$,
where $a$ is the lattice spacing.

Upon expanding the quadratic term appearing in $\mathcal{L}_{I}$, we
get two sorts of four-fermion couplings
\begin{eqnarray}
\mathcal{L}_{I} = g\sum_{\sigma}\bar{\Psi}_{\sigma}\Psi_{\sigma}
\bar{\Psi}_{\sigma}\Psi_{\sigma} - 2g\bar{\Psi}_{\sigma_1}
\Psi_{\sigma_1} \bar{\Psi}_{\sigma_2} \Psi_{\sigma_2}.
\end{eqnarray}
For a given spin $\sigma$, the coupling
$\bar{\Psi}_{\sigma}\Psi_{\sigma} \bar{\Psi}_{\sigma}\Psi_{\sigma}$
amounts to the GN interaction with flavor $N=1$. Such a coupling
term cannot be treated by means of $1/N$ expansion. The coupling
constant $g$ has the dimension of inverse mass. It is convenient to
define a dimensionless parameter $\tilde{g} = g\Lambda/v$, where the
momentum cutoff $\Lambda$ is connected to $a$
via the relation $\Lambda \sim a^{-1}$. In the following, we will
choose to carry out series expansion in powers of $\tilde{g}$. This
method is invalid in the strong coupling regime. Tang \emph{et al.}
\cite{Tang18} have numerically investigated the strong coupling
regime by means of QMC simulation and found that the system
enters into AFM Mott insulating phase once $|\tilde{g}|$ becomes
sufficiently large.

\begin{figure}
\includegraphics[width=0.49\textwidth]{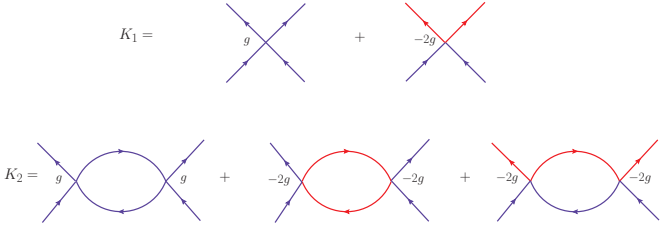}
\caption{Diagrams of leading ($K_1$) and sub-leading ($K_2$) order
contributions to the GN interaction kernel. Blue and red solid lines
stand for fermions with spin $\sigma$ and spin $-\sigma$,
respectively.} \label{Fig:Fourfermion-Kernel}
\end{figure}

We first ignore the Coulomb interaction and examine whether or not
the pure spinful GN interaction leads to dynamical generation of
excitonic gap. According to the analysis of Ref.~\cite{Dorey90}, the
DS equation can be formally written as
\begin{eqnarray}
G^{-1}_{\sigma} &=& (G^{0}_{\sigma})^{-1} - \sum_{\sigma'}
\mathrm{Tr}\left[K_{\sigma,\sigma'}G_{\sigma}\right] \nonumber \\
&& +\frac{1}{2} \sum_{\sigma'} \mathrm{Tr}\left[G_{\sigma'}
\frac{\delta K_{\sigma,\sigma'}}{\delta
G_{\sigma}}G_{\sigma'}\right],
\end{eqnarray}
where $K_{\sigma,\sigma'}$ is the four-fermion interaction kernel.
$K_{\sigma,\sigma'}$ can be obtained from the sum of all the
two-particle irreducible vacuum diagrams in the full fermion
propagators \cite{Dorey90}, represented by $V_{2\mathrm{IR}}(G)$.
$V_{2\mathrm{IR}}(G)$ is connected to the kernel $K_{\sigma\sigma'}$
in the following way
\begin{eqnarray}
V_{2\mathrm{IR}}(G) = \sum_{\sigma,\sigma'}
\frac{1}{2}\mathrm{Tr}[G_{\sigma} K_{\sigma,\sigma'}G_{\sigma'}].
\end{eqnarray}
In this paper, we will retain both the leading order and sub-leading
order corrections. The corresponding Feynman diagrams are presented
in Fig.~\ref{Fig:Fourfermion-Kernel}.

The leading order contributions to $K$ are
\begin{eqnarray}
(K_{1})_{\sigma_{1},\sigma_{1}} = \frac{v\tilde{g}}{\Lambda}, \quad
(K_{1})_{\sigma_1,\sigma_{2}} = -2\frac{v\tilde{g}}{\Lambda}.
\end{eqnarray}
The sub-leading order contributions are
\begin{eqnarray}
(K_{2})_{\sigma_{1},\sigma_{1}}(q) &=& -\int \frac{d^{3} k}{(2
\pi)^{3}}\mathrm{Tr}\left[\left(\frac{v\tilde{g}}{\Lambda}\right)^{2}
G^{0}_{\sigma_{1}}(k) G^{0}_{\sigma_{1}}(q+k)\right]\nonumber \\
&& -\int\frac{d^{3}k}{(2\pi)^{3}} \mathrm{Tr}
\left[\left(2\frac{v\tilde{g}}{\Lambda}\right)^{2}
G^{0}_{\sigma_{2}}(k)G^{0}_{\sigma_{2}}(q+k)\right]\nonumber \\
&=& -5\left(\frac{v\tilde{g}}{\Lambda}\right)^{2}\Pi_{g}(q), \\
(K_{2})_{\sigma_{1},\sigma_{2}}(q) &=&
-\int\frac{d^{3}k}{(2\pi)^{3}}\mathrm{Tr}\left[(2
\frac{v\tilde{g}}{\Lambda})^{2}G^{0}_{\sigma_{1}}(k)
G^{0}_{\sigma_{2}}(q+k)\right] \nonumber \\
&=& -4\left(\frac{v\tilde{g}}{\Lambda}\right)^{2}\Pi_{g}(q),
\end{eqnarray}
where $q \equiv (\Omega,\mathbf{q})$ and $k \equiv
(\omega,\mathbf{k})$, and we define
\begin{eqnarray}
\Pi_{g}(q) = \int\frac{d\omega}{2\pi}
\frac{d^{2}\mathbf{k}}{(2\pi)^2}{(2\pi)^{3}}
\mathrm{Tr}\left[G^{0}_{\sigma_{i}}(k)
G^{0}_{\sigma_{i}}(q+k)\right],
\end{eqnarray}
which is independent of spin directions. After doing simple
calculations, we find that $\Pi_{g}(q) = \frac{1}{4v^{2}}
\sqrt{v^{2}\mathbf{q}^{2} - \Omega^{2}}$. From
Fig.~\ref{Fig:Fourfermion-Kernel}, we see that the first two orders
of corrections satisfy the relation \cite{Dorey90} $\frac{\delta
K_{i}}{\delta G} = 0$ for both $i=1$ and $i=2$. Therefore, the DS
equation for fermion self-energy takes the form
\begin{widetext}
\begin{eqnarray}
i\Sigma_{\sigma}(\varepsilon,\mathbf{p}) &=& -\sum_{\sigma'}\int
\frac{d\omega}{2\pi} \frac{d^{2}\mathbf{k}}{(2\pi)^2}\mathrm{Tr}
\left[i(K_{1})_{\sigma,\sigma'}(\Omega,\mathbf{q})
iG_{\sigma'}(\omega,\mathbf{k})\right] +
\sum_{\sigma'}\int\frac{d\omega}{2\pi} \frac{d^{2}\mathbf{k}}{(2
\pi)^2}(i(K_{2})_{\sigma,\sigma'}(\Omega,\mathbf{q}))(i
G_{\sigma'}(\omega,\mathbf{k})) \nonumber \\
&=&-\frac{v\tilde{g}}{\Lambda}\int\frac{d\omega}{2\pi} \frac{d^{2}
\mathbf{k}}{(2\pi)^2}\mathrm{Tr}[G_{\sigma}(\omega,\mathbf{k})] +
9\left(\frac{v\tilde{g}}{\Lambda}\right)^{2} \int
\frac{d\omega}{2\pi} \frac{d^{2}\mathbf{k}}{(2\pi)^2}{(2\pi)^{3}}
\Pi_{g}(\Omega,\mathbf{q}) G_{\sigma}(\omega,\mathbf{k}).
\label{Eq:FourfermionSelfenergyE}
\end{eqnarray}
In the small $\tilde{g}$ region, we ignore the fermion damping and
velocity renormalization, thus the fermion self-energy can be
identified as the excitonic mass gap. We derive the following DS gap
equation
\begin{eqnarray}
m_{\sigma}(\varepsilon,\mathbf{p}) &=&
i\frac{4v\tilde{g}}{\Lambda}\int\frac{d\omega}{2\pi} \frac{d^{2}
\mathbf{k}}{(2\pi)^2} \frac{m_{\sigma}(\omega,
\mathbf{k})}{\omega^{2}-v^{2}\mathbf{k}^{2} - m_{\sigma}^{2}(\omega,
\mathbf{k})} - i9\left(\frac{v\tilde{g}}{\Lambda}\right)^{2}\int
\frac{d\omega}{2\pi} \frac{d^{2}\mathbf{k}}{(2\pi)^2}
\Pi_{g}(\Omega,\mathbf{q}) \frac{m_{\sigma}(\omega,
\mathbf{k})}{\omega^{2}-v^{2}\mathbf{k}^{2} -
m_{\sigma}^{2}(\omega,\mathbf{k})}. \nonumber
\label{Eq:FourfermionGapEquationM}
\end{eqnarray}
After Wick rotation, this equation is re-cast as
\begin{eqnarray}
m_{\sigma}(\varepsilon,\mathbf{p}) &=& \frac{4v\tilde{g}}{\Lambda}
\int\frac{d\omega}{2\pi} \frac{d^{2}\mathbf{k}}{(2\pi)^2}
\frac{m_{\sigma}(\omega,\mathbf{k})}{\omega^{2} + v^{2}\mathbf{k}^{2} +
m_{\sigma}^{2}(\omega,\mathbf{k})} - \frac{9}{4v^{2}}
\left(\frac{v\tilde{g}}{\Lambda}\right)^{2} \int\frac{d\omega}{2\pi}
\frac{d^{2}\mathbf{k}}{(2\pi)^2}\frac{m_{\sigma}(\omega,\mathbf{k})
\sqrt{v^{2}\mathbf{q}^{2} + \Omega^{2}}}{\omega^{2} + v^{2}\mathbf{k}^{2}
+ m_{\sigma}^{2}(\omega,\mathbf{k})}.
\label{Eq:ShortrangegapEquation}
\end{eqnarray}
\end{widetext}

The leading order correction to dynamical gap generation has been
previously analyzed in Ref.~\cite{Kaveh05}. Notice there is a sign
difference in the definition of $\tilde{g}$. In Ref.~\cite{Kaveh05},
a finite gap is generated only when $\tilde{g} < \tilde{g}_{c} =
-\pi^{2}/4$ (in the limit of $N\rightarrow \infty$); in our case the
critical value becomes $\tilde{g}_{c} = \pi^{2}/2$. In this work we
only consider negative $\tilde{g}$, thus the GN interaction cannot
induce excitonic pairing by itself. However, the GN interaction
might affect the fate of excitonic pairing induced by the Coulomb
interaction. This will be studied in
Sec.~\ref{Sec:OnsiteNumResults}.

\subsection{Yukawa coupling interaction near AFM QCP}

When the strength of spinful GN interaction increases, the AFM
correlation is enhanced. The gapless Dirac SM becomes an AFM Mott
insulator once $U$ exceeds some critical value $U_c$, which defines
a zero temperature AFM QCP. As revealed by Tang \emph{et al.}
\cite{Tang18}, $U_c$ appears to be an increasing function of
$\alpha$ in the region of weak Coulomb interaction. Previous studies
on such an AFM quantum criticality \cite{Herbut06, Herbut09}
demonstrated that the Yukawa coupling between Dirac fermion and AFM
quantum fluctuation, described by scalar field $\phi$, determines
the low-energy properties of the AFM QCP if the Coulomb interaction
is ignored. Here, we are particularly interested in whether the
excitonic pairing is suppressed or promoted near the AFM QCP.

To describe the AFM fluctuation, we add to $\mathcal{L}_{0}$ the
following Lagrangian density of $\phi$ field
\begin{eqnarray}
\mathcal{L}_{b} = -\phi(\partial_{\tau}^{2} + v_{\phi}^{2}
\mathbf{\bigtriangledown}^{2} + r)\phi -
\frac{\lambda_{0}}{4!}\phi^{4} +\sum_{\sigma}\lambda\phi\cdot \sigma
\bar{\Psi}_{\sigma}\Psi_{\sigma}, \nonumber \\
\end{eqnarray}
where $\lambda$ is the strength parameter for Yukawa coupling
interaction and $\sigma = \pm 1$ is fermion spin. The AFM order
parameter \cite{Herbut06, Tang18} is given by $A = \langle
\sum_{\sigma}\sigma \bar{\Psi}_{\sigma}\Psi_{\sigma}\rangle$. The
scalar field $\phi$ stands for the quantum fluctuation around this
mean value. The boson mass $r$ can be identified as the tuning
parameter for SM-AFM transition, and $r = 0$ at the QCP. Here, we
only consider the SM side of the QCP, and thus suppose $r \geq 0$.
To facilitate analytical calculations, we introduce two new coupling
constants for two spin components: $\lambda_{\sigma} = \lambda
\sigma$. It is worth mentioning that $\lambda_{\sigma}$ have the
same dimension as $\sqrt{r}$.

\begin{figure}
\includegraphics[width=0.5\textwidth]{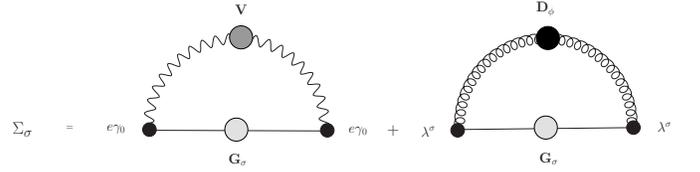}
\caption{Feynman diagram of the fermion self-energy due to Coulomb
interaction and Yukawa coupling.} \label{Fig:selfenergy}
\end{figure}

Once $\mathcal{L}_{b}$ is introduced, the continual chiral symmetry
is explicitly broken. Nevertheless, the total action still preserves
a discrete chiral symmetry $\Psi_{\sigma}\rightarrow
\gamma_{5}\Psi_{\sigma}$, so long as the scalar field $\phi$
transforms simultaneously in the following way: $\phi \rightarrow
-\phi$. When a finite gap is generated via excitonic pairing, the
above discrete chiral symmetry will be dynamically broken.

The Feynman diagrams of the fermion self-energy are shown in
Fig.~\ref{Fig:selfenergy}, where $V$ stands for the dressed Coulomb
interaction function and $D_{\phi}$ for the dressed propagator of
the $\phi$ field. To the leading order of $1/N$ expansion, the
dynamical gap satisfies the following DS equation
\begin{eqnarray}
m_{\sigma}(\varepsilon,\mathbf{p}) &=& \frac{i}{4}\int
\frac{d\omega}{2\pi}\frac{d^2\mathbf{k}}{(2\pi)^2}
\textmd{Tr}[\gamma_{0}G_{\sigma}(\omega,\mathbf{k})
V(\Omega,\mathbf{q})\gamma_{0}] \nonumber \\
&& +\frac{i(\lambda_{\sigma})^{2}}{4}\int\frac{d\omega}{2\pi}
\frac{d^2\mathbf{k}}{(2\pi)^2}\textmd{Tr}[G_{\sigma}(\omega,\mathbf{k})
D_{\phi}(\Omega,\mathbf{q})], \nonumber \\
\label{Eq:FermionSelfenergy}
\end{eqnarray}
where $\varepsilon = \Omega+\omega$ and $\mathbf{p} =
\mathbf{q}+\mathbf{k}$.

The free propagator of the bosonic field $\phi$ is
$$D_{\phi}^{0}(\Omega,\mathbf{q}) = \frac{1}{\Omega^{2} -
v^{2}\mathbf{q}^{2} - r^{2}}.$$ Similar to the long-range Coulomb
interaction, here we assume the boson velocity equals to the fermion
velocity. This free propagator is also renormalized by the
collective excitations. Including this effect leads to the following
dressed bosonic propagator
\begin{eqnarray}
D_{\phi}(\Omega,\mathbf{q}) = \frac{1}{D_{\phi}^{0}
(\Omega,\mathbf{q})^{-1}+\Pi_{\phi}(\Omega,\mathbf{q})},
\end{eqnarray}
where the screening effect is embodied in the polarization function
$\Pi_{\phi}(\Omega,\mathbf{q})$. To the leading order of $1/N$
expansion, the diagram of $\Pi_{\phi}(\Omega,\mathbf{q})$ is
presented in Fig.~\ref{Fig:Coulomb polarization}, given by the
integral
\begin{eqnarray}
\Pi_{\phi}(\Omega,\mathbf{q}) &=& -\sum_{\sigma}\int
\frac{d\omega}{2\pi}\frac{d^2\mathbf{k}}{(2\pi)^2}
\textmd{Tr}[\lambda_{\sigma} G^{0}_{\sigma}(\omega,\mathbf{k})
\lambda_{\sigma}
\nonumber \\
&& \times G^{0}_{\sigma}(\omega+\Omega,\mathbf{k+q})].
\end{eqnarray}
According to the detailed calculations presented in the
Appendix.~\ref{Sec:AppendixPola}, $\Pi_{\phi}$ has the simple form
\begin{eqnarray}
\Pi_{\phi}(\Omega,\mathbf{q}) =
-\frac{N(\lambda_{\sigma})^{2}}{v^{2}}
\frac{\sqrt{v^{2}\mathbf{q}^{2}-\Omega^{2}}}{4}.
\label{Eq:ScalarPolarization}
\end{eqnarray}
which is consistent with that obtained in Ref.~\cite{Lars11}.

After performing calculations, we get the gap equation
\begin{widetext}
\begin{eqnarray}
m_{\sigma}(\varepsilon,\mathbf{p}) &=& \int \frac{d\omega}{2
\pi}\frac{d^2\mathbf{k}}{(2\pi)^2}
\frac{m_{\sigma}(\omega,\mathbf{k}) }{\omega^{2}+v^{2}\mathbf{k}^{2} +
m_{\sigma}(\omega,\mathbf{k})^{2}} \frac{1}{\frac{|\mathbf{q}|}{2\pi
v\alpha} + \frac{N}{8}
\frac{\mathbf{q}^{2}}{\sqrt{\Omega^{2}+v^{2}\mathbf{q}^{2}}}}
\nonumber \\
&& -(\lambda_{\sigma})^{2}\int\frac{d\omega}{2\pi}
\frac{d^2\mathbf{k}}{(2\pi)^2}\frac{m_{\sigma}(\omega,\mathbf{k})
}{\omega^{2}+v^{2}\mathbf{k}^{2} +
m_{\sigma}(\omega,\mathbf{k})^{2}}\frac{1}{\Omega^{2} +
v^{2}\mathbf{q}^{2}+r^{2}+\frac{N(\lambda_{\sigma})^{2}}{v^{2}}
\frac{\sqrt{\Omega^{2}+v^{2}\mathbf{q}^{2}}}{4}},
\label{Eq:FullFermionSelfenergy}
\end{eqnarray}
\end{widetext}
where Wick rotation has been performed. There is a minus sign in the
contribution due to the Yukawa coupling interaction. Two important
conclusions can be deduced. First, the Yukawa coupling tends to
suppress excitonic pairing. Second, the Yukawa coupling by itself is
not able to trigger excitonic pairing.

\section{Numerical results \label{Sec:NumResults}}

In this section, we solve the DS equations numerically and analyze
the physical implications of the solutions. We will first consider
the case of weak GN interaction and then the vicinity of AFM QCP.
Our aim is to determine their influence on the value of $\alpha_c$.
To make numerical evaluation easier, we carry out the following
re-scaling transformations:
\begin{eqnarray}
\frac{m_{\sigma}}{v\Lambda}\rightarrow m_{\sigma},\,\,
\frac{|\mathbf{p}|}{\Lambda}\rightarrow \mathbf{p},\,\,
\frac{|\mathbf{k}|}{\Lambda}\rightarrow \mathbf{k},\,\,
\frac{|\mathbf{q}|}{\Lambda}\rightarrow \mathbf{q},\,\,\nonumber
\\
\frac{\lambda_{\sigma}^{2}}{v\Lambda} \rightarrow \lambda_{\sigma}^{2},\,\,
\frac{\omega}{v\Lambda} \rightarrow \omega,\,\,
\frac{\Omega}{v\Lambda} \rightarrow \Omega,\,\,
\frac{\varepsilon}{v\Lambda} \rightarrow \varepsilon.
\end{eqnarray}
By doing so, all the parameters that appear in the gap equations are
made dimensionless.

\subsection{Interplay of Coulomb and GN interactions
\label{Sec:OnsiteNumResults}}

To examine the interplay between the long-range Coulomb and
short-range GN interactions, we combine
Eq.~\ref{Eq:CoulombgapEquation} and
Eq.~\ref{Eq:ShortrangegapEquation}, and then solve the total gap
equation self-consistently for different values of $\tilde{g}$ at
$N=2$ and $\alpha = 2.2$, which are the physical flavor and the
physical $\alpha$ of suspended graphene.

To simplify numerical evaluation, it is useful to first adopt the
commonly used instantaneous approximation \cite{Khveshchenko01},
which assumes that the fermion gap, the Coulomb interaction
function, and the four-fermion interaction kernel are independent of
energy. The impact of the energy dependence will be examined later.
Under this approximation, the total gap equation can be written as
\begin{eqnarray}
m_{\sigma}(\mathbf{p}) &=& \int \frac{d^2\mathbf{k}}{(2
\pi)^2}\frac{m_{\sigma}(\mathbf{k})}{2\sqrt{\mathbf{k}^{2} +
m_{\sigma}(\mathbf{k})^{2}}} \frac{1}{\frac{|\mathbf{q}|}{2\pi
\alpha} + \frac{N}{8}|\mathbf{q}|}\nonumber \\
&& +\tilde{g}\int \frac{d^{2}\mathbf{k}}{(2\pi)^{2}}
\frac{m_{\sigma}(\mathbf{k})}{2\sqrt{\mathbf{k}^{2} +
m_{\sigma}^{2}(\mathbf{k})}} \nonumber \\
&& -\frac{9}{4} \left(\tilde{g}\right)^2
\int\frac{d^{2}\mathbf{k}}{(2\pi)^{2}}\frac{|\mathbf{q}|
m_{\sigma}(\mathbf{k})}{2\sqrt{\mathbf{k}^{2} +
m_{\sigma}^{2}(\mathbf{k})}}.
\end{eqnarray}

\begin{figure}
\includegraphics[width=0.46\textwidth]{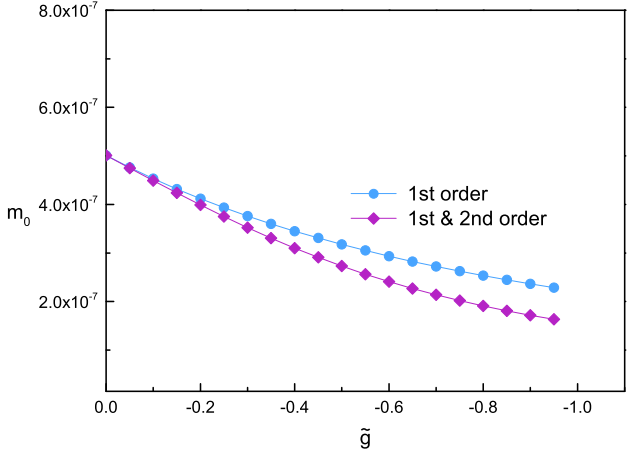}
\caption{The $\tilde{g}$-dependence of zero-momentum excitonic gap
$m_{0}$ at $\alpha=2.2$ and $N=2$. }
\label{Fig:InstantaneousGapFourfermion2.2}
\end{figure}

\begin{figure}
\includegraphics[width=0.44\textwidth]{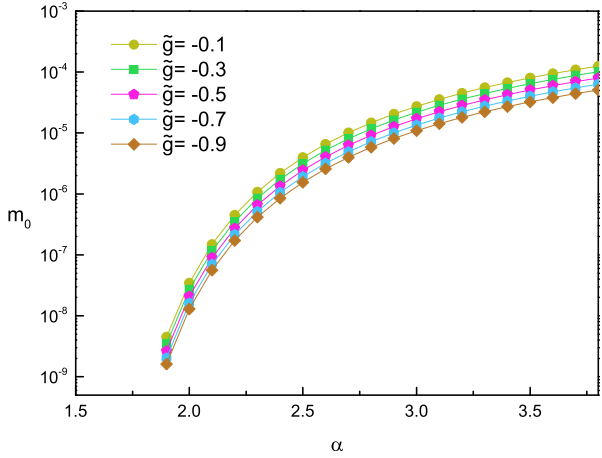}
\caption{The $\alpha$-dependence of zero-momentum gap $m_{0}$ for
different values of $v\tilde{g}$ at $N=2$.}
\label{Fig:InstantaneousGapFourfermionAlpha}
\end{figure}

After solving this equation, we present the numerical results in
Fig.~\ref{Fig:InstantaneousGapFourfermion2.2}, where $m_0$ is
defined as the value of fermion gap at zero
momentum. As $|\tilde{g}|$ grows, $m_0$ decreases considerably. This
implies that weak GN interaction tends to suppress excitonic gap.
The $\alpha$-dependence of $m_0$ is shown in
Fig.~\ref{Fig:InstantaneousGapFourfermionAlpha}. We observe that, as
GN interaction increases, the value of $\alpha_c$ will be slightly
increased. When $\tilde{g} = -0.7$, we find that $\alpha_{c} = 2.0$.

The instantaneous approximation has been previously used in the DS
study of excitonic gap generation \cite{Khveshchenko01}. Extensive
works confirmed that the value of $\alpha_c$ obtained under this
approximation is actually not far from that obtained by
incorporating higher order corrections. In this sense, the
instantaneous approximation leads to qualitatively reliable
conclusion. Further, we have also solved the gap equation
incorporating the energy dependence of gap function and Coulomb
interaction. The sub-leading order correction due to spinful GN
interaction exhibits a logarithmic dependence on the energy cutoff.
Our numerical results show that, increasing the energy cutoff more
or less modify the values of $\tilde{g}_c$ and $\alpha_{c}$.
However, the qualitative conclusion that GN interaction suppresses
excitonic pairing is not changed. Here, we choose an energy cutoff
$\Lambda_{E} = 10v\Lambda$.

\subsection{Interplay of Coulomb interaction and Yukawa coupling interaction near AFM QCP}

We now consider the interplay of Coulomb interaction and Yukawa
coupling interaction. The corresponding DS gap equation is given by
Eq.~\ref{Eq:FullFermionSelfenergy}. Numerical calculations verify
that the solution of this gap equation is insensitive to the energy
cutoff. Below, the energy cutoff is taken as $\Lambda_{E} = 1000
v\Lambda$.

To get a rapid glimpse of the main results, we will first neglect
the energy dependence of both the fermion self-energy and the
interaction functions. This approximation can be implemented by
making the following replacement:
\begin{eqnarray}
m_{\sigma}(\varepsilon,\mathbf{p})&\rightarrow&
m_{\sigma}(\mathbf{p}), \\
V(\Omega,\mathbf{q})&\rightarrow& V(\mathbf{q}),\\
D_{\phi}(\Omega,\mathbf{q})&\rightarrow& D_{\phi}(\mathbf{q}).
\end{eqnarray}
Under this approximation, the DS gap equation becomes
\begin{eqnarray}
m_{\sigma}(\mathbf{p}) &=& \int \frac{d^2\mathbf{k}}{(2\pi)^2}
\frac{m_{\sigma}(\mathbf{k})}{2\sqrt{\mathbf{k}^{2} +
m_{\sigma}(\mathbf{k})^{2}}}\frac{1}{\frac{|\mathbf{q}|}{2\pi
\alpha} + \frac{N}{8}|\mathbf{q}|}\nonumber \\
&& -(\lambda_{\sigma})^{2}\int\frac{d^2\mathbf{k}}{(2\pi)^2}
\frac{m_{\sigma}(\mathbf{k})}{2\sqrt{\mathbf{k}^{2}
+ m_{\sigma}(\mathbf{k})^{2}}} \nonumber \\
&& \times \frac{1}{\mathbf{q}^{2} + r^{2} +
N(\lambda_{\sigma})^{2}\frac{|\mathbf{q}|}{4}}.
\label{Eq:InstantaneousFermionSelfenergy}
\end{eqnarray}

\begin{figure}
\includegraphics[width=0.44\textwidth]{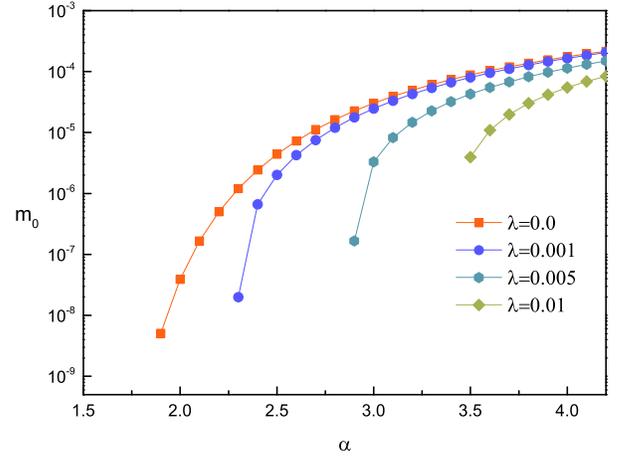}
\caption{The $\alpha$-dependence of $m_{0}$ for different values of
$\lambda$ at $N=2$. Clearly, $\alpha_c$ is an increasing function of
$\lambda$.} \label{Fig:InstantaneousAlphaN2}
\end{figure}

We have solved this equation numerically, and shown in
Fig.~\ref{Fig:InstantaneousAlphaN2} the $\alpha$-dependence of
excitonic gap obtained at zero momenta, namely $m(p=0)$, for
different values of $\lambda$. If the AFM QCP is entirely ignored,
corresponding to $\lambda = 0$, the critical value $\alpha_c \approx
1.9$. If $\lambda$ takes a very small value $\lambda=0.001$,
$\alpha_c$ is increased to $\alpha_c \approx 2.3$. For $\lambda =
0.005$ and $\lambda = 0.01$, we find that $\alpha_{c} = 2.8$ and
$\alpha_c = 3.5$, respectively. Therefore, the excitonic gap
generation can be significantly suppressed at the AFM QCP.

\begin{figure}
\includegraphics[width=0.46\textwidth]{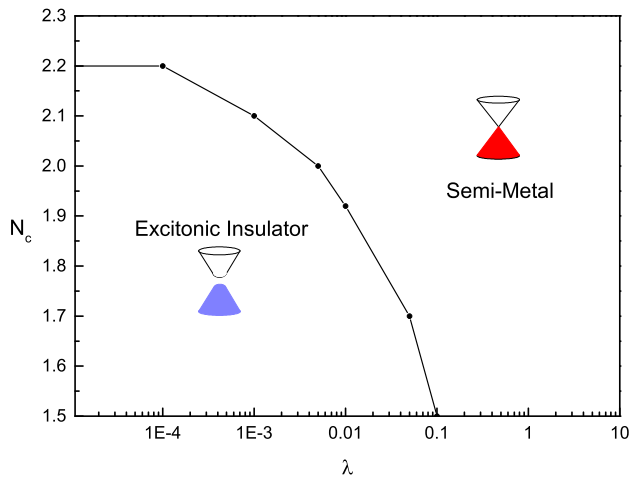}
\caption{The critical line of SM-EI transition on the $\lambda$-$N$
plane. Here the Coulomb interaction parameter is fixed at
$\alpha=3.2$.} \label{Fig:InstantaneousPhasediagramN-g}
\end{figure}

Further, we study how $\lambda$ changes the critical fermion flavor
$N_{c}$. We fix $\alpha$ at $\alpha = 3.2$, and solve Eq.~(34) to
obtain the relation between $\lambda$ and $N_c$, with results
presented in Fig.~\ref{Fig:InstantaneousPhasediagramN-g}. For very
small values of $\lambda$, $N_c \approx 2.2$. For $\lambda > 0.005$,
$N_{c}$ is reduced below $2$. For Dirac fermions with physical
flavor $N=2$, the excitonic pairing cannot occur due to the presence
of sufficiently strong Yukawa coupling interaction.

\begin{figure}
\includegraphics[width=0.45\textwidth]{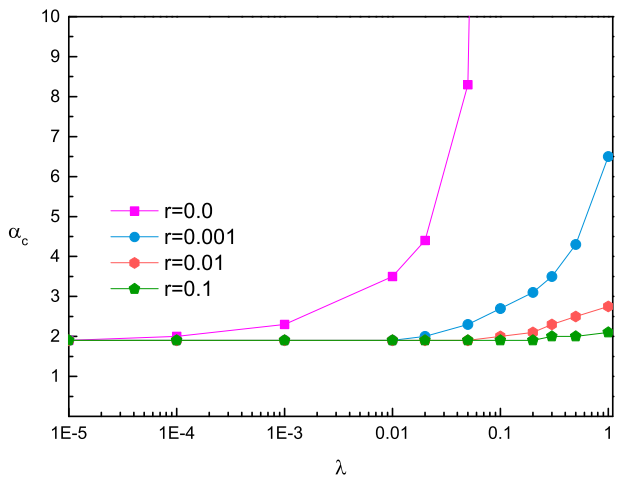}
\caption{The $\lambda$-dependence of $\alpha_c$ at a number of
values of $r$. Here the fermion flavor is $N=2$.}
\label{Fig:InstantaneousPhasediagrama-g}
\end{figure}

The dependence of $\alpha_c$ on $\lambda$ at fixed flavor $N=2$ is
shown in Fig.~\ref{Fig:InstantaneousPhasediagrama-g}. At the AFM QCP
with $r=0$, $\alpha_c$ increases rapidly as $\lambda$ grows, and
finally goes to infinity at sufficiently large $\lambda$. This is
another signature that AFM quantum criticality disfavors excitonic
gap generation. As $r$ increases, the system moves away from the AFM
QCP into the SM region. In this process, the quantum fluctuation of
AFM order parameter is weakened, and the suppressing effect of
excitonic pairing becomes progressively unimportant.

In Ref.~\cite{Tang18}, the authors found that the Coulomb
interaction tends to increase $U_c$. Here our finding is that AFM
quantum fluctuation suppresses excitonic pairing. There seems to be
a repulsion between the excitonic pairing and the AFM ordering.
Based on these results, we plot a schematic phase diagram on the
$U$-$\alpha$ plane in Fig.~\ref{Fig:ExcitonicMottphasediagram}. The
critical line of $U_c$ goes rightwards as $\alpha$ increases
\cite{Tang18}, whereas the critical line of $\alpha_c$ goes upwards
as $U \rightarrow U_c$ from the left side. A previous RG study
\cite{Herbut06} predicted that the Dirac fermion system may undergo
a first-order transition from EI to AFM Mott insulator. Our results
indicate that such a direct transition does not occur, and that the
excitonic insulating phase and AFM Mott insulating phase are
actually separated by an intermediate gapless SM phase.

We have also numerically solved
Eq.~(\ref{Eq:FullFermionSelfenergy}), in which $\alpha_{c}=0.7$ at
$\lambda=0$, and reached the same conclusion that the quantum AFM
criticality tends to suppress excitonic gap generation. Therefore,
the schematic phase diagram presented in Fig.~1 is still
qualitatively correct after taking the energy dependence into
account.

\section{Application of the results}

Determination of the precise value of $\alpha_c$ proves to be a
highly nontrivial challenge. In QED$_{3}$, the fermion propagator
and the gauge boson propagator are coupled to each other by a set of
DS integral equations \cite{Roberts94}. The same is true for our
case, since the Coulomb interaction can be effectively described by
the coupling between the Dirac fermion and the temporal component of
U(1) gauge boson. The full fermion propagator has the following
generic form
\begin{eqnarray}
G(\varepsilon,\mathbf{p}) = \frac{1}{\gamma_{0}\varepsilon
Z(\varepsilon,\mathbf{p}) - v(\gamma_{1}p_{x} +
\gamma_{2}p_{y})A(\varepsilon,\mathbf{p}) -
m(\varepsilon,\mathbf{p})}, \nonumber
\end{eqnarray}
where $Z(\varepsilon,\mathbf{p})$ and $A(\varepsilon,\mathbf{p})$
are the wave function renormalizations and
$m(\varepsilon,\mathbf{p})$ is the fermion gap function. The dressed
boson propagator is given by Eq.~\ref{Eq:CoulombPropagator}, where
the polarization function should be replaced by the full one. The
full polarization is determined by the dressed fermion propagator
and the interaction vertex function $\Gamma$. Once the expression
for vertex function is known, the dressed fermion and boson
propagators could be determined. In practice, it is not possible to
obtain the exact solutions of the coupled DS equations, and one
always needs to introduce certain approximations (truncations) to
replace the full propagators and the full vertex function with
approximate ones. In the literature there are two commonly used
vertex functions: the bare vertex and the Ball-Chiu \cite{BallChiu}
vertex.

\begin{table}[htbp]
\caption{DS equation results for the critical value $\alpha_c$ at
the flavor $N=2$. $Z$, $A$, $m$, and $\alpha$ are defined in the
context. $\Gamma_{\mathrm{BC}}$ stands for the Ball-Chiu vertex
correction. $\Pi_{c}^{\mathrm{RPA}}$ is the RPA-level polarization
given by Eq.~(8), and $\Pi_{c}^{\mathrm{SC}}$ represents the
polarization function obtained from self-consistent DS equation
calculations. The symbol $\otimes$ refers to the instantaneous
approximation, and $\surd$ indicates that the energy dependence is
taken into account. The corresponding function is neglected if the
space is left blank. \label{Table:CriticalInteraction}}
\begin{center}
\begin{tabular}{|C{0.8cm}|C{0.8cm}|C{0.8cm}|C{0.8cm}|C{0.8cm}|C{0.8cm}|C{0.8cm}|C{2.0cm}|}
\hline\hline $Z$ & $A$ & $m$ & $\Gamma_{\mathrm{BC}}$ &
$\Pi_{c}^{\mathrm{SC}}$ & $\Pi_{c}^{\mathrm{RPA}}$ & $\alpha$ &
Reference
\\
\hline  &   & $\otimes$ &   &   & $\otimes$ & 1.9 & current paper
\\
\hline   &   & $\otimes$ &   &   & $\surd$ & 0.92 & \cite{Gamayun09}
\\
\hline  &   & $\surd$ &   &   & $\surd$ & 0.7 & current paper
\\
\hline $\surd$ & $\surd$ & $\surd$ & $\surd$ &   &
$\surd$ & 3.2 & \cite{WangLiu12}
\\
\hline $\surd$ & $\surd$ & $\surd$ & $\surd$ &  &
$\surd$ & 2.9 & \cite{Carrington16}
\\
\hline $\surd$ & $\surd$ & $\surd$ & $\surd$ & $\otimes$ & & 1.99 &
\cite{Carrington18}
\\
\hline $\surd$ & $\surd$ & $\surd$  & $\surd$ &
$\surd$ &   & 2.06 & \cite{Carrington18}
\\
\hline\hline
\end{tabular}
\end{center}
\end{table}

Extensive DS equation studies of excitonic pairing in graphene have
revealed that the precise value of $\alpha_c$ is very sensitive to
the specific approximation. To demonstrate this, we list in Table I
a number of representative values of $\alpha_c$ obtained by
employing various approximations. If one assumes $Z = A = 1$ and
ignores the vertex correction, the critical value $\alpha_c = 1.9$
in the instantaneous approximation and $\alpha_c = 0.7$ after
including the energy dependence. Two of the authors \cite{WangLiu12}
have incorporated $Z$ and $A$, utilized the first term of Ball-Chiu
vertex, and adopted the RPA expression of polarization $\Pi_c$.
Under such approximations, it was found \cite{WangLiu12} that
$\alpha \approx 3.2$. It was pointed out in Ref.~\cite{WangLiu12}
that $\alpha_c$ could be considerably decreased if the feedback of
excitonic gap on $\Pi_c$ is included. Recently, Carrington \emph{et
al.} \cite{Carrington18} have carried out more refined DS equation
calculations after taking into account $Z$, $A$, the first term of
Ball-Chiu vertex, and also the feedback effects of $Z$, $A$, and $m$
on $\Pi_c$, and obtained $\alpha_c \approx 2.06$. It is surprising
that the value $\alpha_c = 1.9$ obtained in the present paper by
employing the crude instantaneous approximation is actually quite
close to the above result. Moreover, the value $\alpha_c$ obtained
by Carrington \emph{et al.} \cite{Carrington18} is smaller than the
physical value $\alpha = 2.16$ of suspended graphene. In the light
of this result, one might have to conclude that clean, undoped
suspended graphene is an EI at low temperatures.

Ellias \emph{et al.} \cite{Geim11} has measured the cyclotron mass
in suspended graphene and found no evidence of finite gap at rather
low energies ($\sim 0.1\mathrm{meV}$). This finding was further
supported by the measurements of Mayorov \emph{et al.}
\cite{Mayorov12}. How can one reconcile the recent theoretical
result of Carrington \emph{et al.} \cite{Carrington18} and these
experiments?

Here we propose that the seeming discrepancy can be explained by
noticing the fact that graphene is not far from the AFM QCP. Wehling
\emph{et al.} \cite{Wehling11} has calculated the on-site
interaction parameter $U$ in suspended graphene by using three
different approaches. The critical value $U_c$ needed to trigger AFM
Mott transition seems to be only slightly larger than the physical
$U$ \cite{Tang18}, which implies that realistic graphene is close to
the AFM QCP \cite{Tang18}. Apparently, the AFM quantum fluctuation
is important in graphene and should be seriously considered in the
study of EI transition. As revealed in our calculations, AFM quantum
fluctuation can strongly suppress excitonic pairing by increasing
the value of $\alpha_c$. Therefore, we conclude that the gapless SM
state of suspended graphene is actually quite robust.

\section{Summary and Discussion \label{Sec:Summary}}

In summary, we have investigated the non-perturbative effect of
dynamical excitonic gap generation in a 2D Dirac fermion system. The
Dirac fermions are subjected to two types of interactions, namely
the long-range Coulomb interaction and the short-range on-site
interaction. The former interaction can trigger excitonic pairing,
whereas the latter leads to AFM Mott insulating quantum phase
transition in the strong coupling regime. The DS equation approach
is employed to study the influence of on-site interaction on the
fate of excitonic gap generation. We first have shown that the
critical Coulomb interaction strength $\alpha_{c}$ is slightly
suppressed by the weak GN interaction. As the system approaches to
the AFM QCP, the dynamics of Dirac fermions is strongly influenced
by the quantum critical fluctuation of AFM order parameter. We have
demonstrated that excitonic gap generation is suppressed by the AFM
quantum fluctuation. Such a suppression effect is most significant
at the AFM QCP, but gradually diminishes when the system moves away
from the QCP. If 2D Dirac fermion system is close to the AFM QCP, as
what happens in graphene, it would be very difficult to generate a
finite excitonic gap.

Based on these results, we provide supplementary information to the
global phase diagram reported in Ref.~\cite{Tang18}. On the phase
diagram, the EI phase is not neighboring to the AFM phase, but is
separated from the AFM phase by an intermediate gapless SM phase, as
illustrated schematically in
Fig.~\ref{Fig:ExcitonicMottphasediagram}. This conclusion is
different from the one previously stated in Ref.~\cite{Herbut06}. As
indicated by our results, it is hardly possible to transform 2D
Dirac fermion system from an EI phase directly to AFM Mott
insulating phase. The reason is that, the quantum critical AFM
fluctuation can effectively prevent excitonic pairing.

When both $\alpha$ and $U$ take large values, the Dirac fermion
system could either be a AFM Mott insulator or a CDW. It might still
be a gapless SM. To determine the quantitatively more precise phase
diagram in such a strongly interacting regime, it is necessary to
investigate the mutual influence between strong Coulomb interaction
and strong on-site interaction in a more self-consistent manner,
which will be carried out in future works.\\

\acknowledgments

The numerical calculations were mainly performed on the
supercomputing system of the Supercomputing Center of the University
of Science and Technology of China. The authors acknowledge the
financial support by the National Natural Science Foundation of
China under Grants No. 11847234, No. 11574285, and No. 11504379, and
the Anhui Provincial Natural Science Foundation under Grants No.
1908085QA16.

\appendix

\begin{widetext}

\section{Calculation of the polarization $\Pi_{\phi}$ \label{Sec:AppendixPola}}

We now provide the calculational details of the polarization
function for the dressed propagator of bosonic AFM fluctuation. To
the leading order of $1/N$ expansion, this polarization is defined
as
\begin{eqnarray}
i\Pi_{\phi}(\Omega,\mathbf{q}) &=& -\sum_{\sigma}\int
\frac{d\omega}{2\pi} \frac{d^{2}\mathbf{k}}{(2\pi)^2}
\textmd{Tr}[\lambda_{\sigma}G^{0}_{\sigma}(\omega,\mathbf{k})\lambda_{\sigma}
G^{0}_{\sigma}(\omega+\Omega,\mathbf{k+q})] \nonumber \\
&=& -\sum_{\sigma}\int\frac{d\omega}{2\pi}
\frac{d^{2}\mathbf{k}}{(2\pi)^{2}}\textmd{Tr}\left[\lambda_{\sigma}
\frac{1}{-\gamma_{0}\omega + v\mathbf{\gamma}\mathbf{k} +
m_{e}}\lambda_{\sigma}\frac{1}{-\gamma_{0}(\omega + \Omega)+
v\mathbf{\gamma}(\mathbf{k+q})+m_{e}}\right],
\end{eqnarray}
where $m_e$ is a constant mass of Dirac fermion. Making the
replacements $\mathbf{q} = v\mathbf{q}$ and $\mathbf{k} =
v\mathbf{k}$, we re-write it in the form
\begin{eqnarray}
i\Pi_{\phi}(\Omega,\mathbf{q}) &=& -\sum_{\sigma}\int
\frac{d\omega}{2\pi}\frac{d^{2}\mathbf{k}}{(2\pi)^2}
\textmd{Tr}\left[\frac{\lambda_{\sigma}}{v^{2}}\frac{1}{-\gamma_{0}
\omega + \mathbf{\gamma}\mathbf{k}+m_{e}}\lambda_{\sigma}
\frac{1}{-\gamma_{0}(\omega+\Omega) + \mathbf{\gamma}(\mathbf{k+q})
+ m_{e}}\right]\nonumber \\
&=&-4\sum_{\sigma}\left(\frac{\lambda_{\sigma}}{v}\right)^{2}\int
\frac{d\omega}{2\pi} \frac{d^{2}\mathbf{k}}{(2\pi)^2}
\frac{((\Omega+\omega)\omega - (\mathbf{k+q})\cdot
\mathbf{k}+m_{e}^{2})}{((\Omega+\omega)^{2} - \mathbf{(k+q)}^{2} -
m_{e}^{2})(\omega^{2} - \mathbf{k}^{2} - m_{e}^{2})}.
\end{eqnarray}
Making use of the Feynman integral
\begin{eqnarray}
\frac{1}{AB} = \int_{0}^{1}dx\frac{1}{[(1-x)A+xB]^{2}},
\end{eqnarray}
we proceed as follows
\begin{eqnarray}
i\Pi_{\phi} &=& -4\sum_{\sigma}\left(\frac{\lambda_{\sigma}}{v}
\right)^{2} \int_{0}^{1} dx \int\frac{d\omega}{2\pi}
\frac{d^{2}\mathbf{k}}{(2\pi)^2} \frac{((\Omega+\omega)\omega -
(\mathbf{k+q})\cdot\mathbf{k} +
m_{e}^{2})}{(x(\Omega+\omega)^{2}-x\mathbf{(k+q)}^{2} -
xm_{e}^{2}+(1-x)\omega^{2}-(1-x)\mathbf{k}^{2} -
(1-x) m_{e}^{2})^{2}}\nonumber \\
&=& -4\sum_{\sigma}\left(\frac{\lambda_{\sigma}}{v}\right)^{2}
\int_{0}^{1}dx \int\frac{d\omega}{2\pi} \frac{d^{2}
\mathbf{k}}{(2\pi)^{2}} \frac{((\Omega+\omega)\omega -
(\mathbf{k+q})\cdot\mathbf{k} + m_{e}^{2})^{2}}{((x -
x^{2})(\Omega^{2}-\mathbf{q}^{2}) + (\omega+x\Omega)^{2} -
(\mathbf{k}+x\mathbf{q})^{2}-m_{e}^{2})^{2}}.
\end{eqnarray}
Define $\omega'=\omega+x\Omega$ and $\mathbf{k'} =
\mathbf{k}+x\mathbf{q}$, we further get
\begin{eqnarray}
i\Pi_{\phi} &=& -4\sum_{\sigma}\left(\frac{\lambda_{\sigma}}{v}
\right)^{2} \int_{0}^{1}dx \int\frac{d\omega'}{2\pi}
\frac{d^{2}\mathbf{k'}}{(2\pi)^2} \frac{(\omega'-x\Omega)(\omega' +
(1-x)\Omega)-(\mathbf{k}'+(1-x)\mathbf{q})\cdot(\mathbf{k}'-x\mathbf{q})
+ m_{e}^{2}}{((x-x^{2})(\Omega^{2}-\mathbf{q}^{2}) + (\omega')^{2} -
{\mathbf{k'}}^{2} - m_{e}^{2})^{2}} \nonumber
\\
&=& -4\sum_{\sigma}\left(\frac{\lambda_{\sigma}}{v}\right)^{2}
\int_{0}^{1}dx \int \frac{d\omega'}{2\pi}
\frac{d^{2}\mathbf{k'}}{(2\pi)^2} \frac{\omega'^{2} +
(1-2x)\Omega\omega'-x(1-x)\Omega^{2}-\mathbf{k'}^{2} -
(1-2x)\mathbf{q\cdot k'}+x(1-x)\mathbf{q}^{2}+m_{e}^{2}}{((x-x^{2})
(\Omega^{2}-\mathbf{q}^{2})+\omega'^{2} -
\mathbf{k'}^{2}-m_{e}^{2})^{2}}. \nonumber \\
\end{eqnarray}
Introducing $C = \sqrt{ (x-x^{2})(\Omega^{2}-\mathbf{q}^{2})-
\mathbf{k'}^{2} - m_{e}^{2}}$ leads to
\begin{eqnarray}
i\Pi_{\phi} = 4\sum_{\sigma}\left(\frac{\lambda_{\sigma}}{v}
\right)^{2} \int_{0}^{1}dx \int\frac{d\omega'}{2\pi}
\frac{d^{2}\mathbf{k'}}{(2\pi)^2}
\left[\frac{\omega'^{2}}{(\omega'^{2}+C^{2})^{2}} -
\frac{(x-x^{2})(\Omega^{2}-\mathbf{q}^{2})+
\mathbf{k'}^{2}-m_{e}^{2}}{(\omega'^{2}+C^{2})^{2}}\right].\nonumber
\end{eqnarray}
Since
\begin{eqnarray}
\int_{-\infty}^{+\infty}dx\frac{x^{2}}{(x^{2} +
a^{2})^{2}}=\frac{\pi}{2a}, \quad \int_{-\infty}^{+\infty}dx
\frac{1}{(x^{2}+a^{2})^{2}}=\frac{\pi}{2a^{3}},
\end{eqnarray}
we find that
\begin{eqnarray}
i\Pi_{\phi} &=& -4\sum_{\sigma}\left(\frac{\lambda_{\sigma}}{v}
\right)^{2}\int_{0}^{1} dx\int \frac{d^{2}\mathbf{k'}}{(2\pi)^3}
\left[\frac{\pi}{2C} - \frac{\pi[C^{2} +
2\mathbf{k'}^{2}]}{2C^{3}}\right] \nonumber \\
&=& 4\sum_{\sigma}\left(\frac{\lambda_{\sigma}}{v}\right)^{2}
\int_{0}^{1}dx\int \frac{d^{2}\mathbf{k'}}{(2\pi)^3}
\frac{2\pi\mathbf{k'}^{2}} {2\sqrt{((x-x^{2})(\Omega^{2} -
\mathbf{q}^{2})-\mathbf{k'}^{2}-m_{e}^{2})^{3}}}.\nonumber
\end{eqnarray}
After carrying out a series of calculations, we eventually obtain
\begin{eqnarray}
i\Pi_{\phi} = -\frac{iN(\lambda_{\sigma})^{2}}{8v^{2}}
\left[\Lambda - 2\sqrt{\mathbf{q}^{2}-\Omega^{2}-m_{e}^{2}}\right],
\end{eqnarray}
where $\Lambda$ is the ultra-violet momentum cutoff. In the massless
limit, i.e., $m_{e} = 0$, we have
\begin{eqnarray}
\Pi_{\phi} = -\frac{N(\lambda_{\sigma})^{2}}{4v^{2}}
\sqrt{v^{2}\mathbf{q}^{2}-\Omega^{2}}.
\end{eqnarray}
After Wick rotation($\Omega\rightarrow i\Omega$), we can have the
polarization function in Euclidean space
\begin{eqnarray}
\Pi_{\phi}^{E} = -\frac{N(\lambda_{\sigma})^{2}}{4v^{2}}
\sqrt{v^{2}\mathbf{q}^{2}+\Omega^{2}}.
\end{eqnarray}
\end{widetext}

\end{document}